\documentclass[aps,pra,twocolumn]{revtex4}
\usepackage{graphicx}
\usepackage{dcolumn}
\usepackage{bm}
\usepackage{amsmath}
\usepackage{amssymb,amsthm}
\usepackage{epsfig,color}
\usepackage[sans]{dsfont}

\definecolor{nblue}{rgb}{0.2,0.2,0.7}
\definecolor{ngreen}{rgb}{0.2,0.6,0.2}
\definecolor{nred}{rgb}{0.7,0.2,0.2}
\definecolor{nblack}{rgb}{0,0,0}

\def\bea{\begin{eqnarray}}
\def\eea{\end{eqnarray}}
\begin{document}
\title{Communication channel of fermionic system in accelerated frame}
\author{Jinho Chang}
\author{Younghun Kwon}

\email{yyhkwon@hanyang.ac.kr}

\affiliation{Department of Physics, Hanyang University,
Ansan, Kyunggi-Do, 425-791, South Korea }

\date{\today}

\begin{abstract}
In this article, we investigate the communication channel of
fermionic system in an accelerated frame. We observe that at the
infinite acceleration, the mutual information of single rail quantum
channel coincides with that of double rail quantum channel, but
those of classical ones reach different values. Furthermore, we find
that at the infinite acceleration, the conditional entropy of
single(or double) rail quantum channel vanishes, but those of
classical ones may have finite values. In addition, we see that even
when considering a method beyond the single mode approximation, the
dual rail entangled state seems to provide better channel capacity
than the single rail entangled state, unlike the bosonic case.
Moreover, we find that the single-mode approximation is not
sufficient to analyze the communication channel of fermionic system
in an accelerated frame.
\end{abstract}

\maketitle

\section*{I.Introduction}

   The quantum communication locates in the center of research about quantum information theory.
 The quantum communication is to find a way to communicate a quantum information between partners.
 The effectiveness of the quantum communication is usually measured by some suitable measures, i.e. mutual information and conditional entropy, etc.\\
 Along the line, the quantum communication in relativistic quantum information is to understand the effectiveness of quantum communication when parties that may share the quantum channels with partner in an inertial frame moves in an accelerated frames\cite{ref:alsing1}\cite{ref:ball}\cite{ref:fuentes}\cite{ref:alsing2}\cite{ref:bruschi}\cite{ref:martin}.
 Actually, in view of the entanglement behavior of relativistic quantum information, there exists the striking effect that even though the entanglement in bosonic system disappears in the infinite acceleration, the entanglement in fermionic system may persist even in the infinite acceleration\cite{ref:fuentes}.\\
 Recently the studies for communication channel of bosonic system in an accelerated frame have been understood\cite{ref:Hosler}\cite{ref:montero3}, where the single and double rail encodings were introduced.
 That is, through a bosonic channel, they try to figure out what is the best method to communicate the quantum information between one in an inertial frame and the other in an accelerated frame.\\
 Likewise, it is important to understand the communication channel of fermionic system specially when one of the partners is in an accelerated frame.
 Up to our knowledge, the study for the channel of the fermionic system was performed only by the single mode approximation which seems not to be a perfect
 one. In fact, a physical ordering beyond single mode approximation\cite{ref:montero1}\cite{ref:montero2}\cite{ref:chang} was recently proposed.
 So using the physical structure,  we investigate the quantum(classical) communication channel such as $\Phi^{+}$and $\Phi^{*}$($\rho^{s},\rho^{d}$), when one partner moves in a uniformly accelerated frame.
 We find out that at the
infinite acceleration, the mutual information of single rail quantum
channel coincides with that of double rail quantum channel even
though those of classical ones reach different values. Furthermore,
we observe that at the infinite acceleration, the conditional
entropy of both single and double rail quantum channel vanishes,
nevertheless those of classical ones may have finite values. In
addition, we can see that even when considering a method beyond the
single mode approximation, the dual rail entangled state seems to
provide better channel capacity than the single rail entangled
state, unlike the bosonic case. Moreover, we note that the
single-mode approximation is not sufficient to analyze the
communication channel of fermionic system
in an accelerated frame.\\
 The organization of this article is as follows. In section II, we will briefly introduce how to describe the fermionic system in a non-inertial frame. In section III we will investigate the channel capacity of fermionic system in an accelerated frame. In section IV we will conclude and discuss our results.

\section*{II. Accelerated Frame}

 The accelerated frame can be described by the Rindler coordinate
 $(\tau,\varsigma,y,z)$ instead of Minkowski coordinate $(t,x,y,z)$,
 having a relation with
\begin{equation}
ct=\varsigma sinh(\frac{a \tau}{c}),x=\varsigma cosh(\frac{a
\tau}{c})
\end{equation}
 Here $a$ denotes the fixed acceleration of the
frame and $c$ is the velocity of light. Eq(1) only covers the region
(I) in right edge. The region (II)(the left edge) can be covered by
$ ct=-\varsigma sinh(\frac{a \tau}{c}),x=-\varsigma cosh(\frac{a
\tau}{c})$.\\
 The field in Minkowski and Rindler spacetime can be written as
\begin{eqnarray}
\phi &=& N_{M}\sum_{i}(a_{i,M}v^{+}_{i,M} +
b^{\dag}_{i,M}v^{-}_{i,M} )\nonumber\\
     &=& N_{R}\sum_{j}(a_{j,I}v^{+}_{j,I} +
b^{\dag}_{j,I}v^{-}_{j,I} + a_{j,II}v^{+}_{j,II} +
b^{\dag}_{j,II}v^{-}_{j,II} ) \nonumber\\
     & &
\end{eqnarray}
 Here $N_{M}$ and $N_{R}$ denote the normalization constants. Also
 $v^{\pm}_{i,M}$($v^{\pm}_{i,I}$
 and $v^{\pm}_{i,II}$) mean(s) the positive and negative energy solutions of the
 Dirac equation in Minkowski spacetime(Rindler spacetime), which can be obtained with
 respect to the Killing vector field in Minkowski spacetime(region I and II). And  $a^{\dag}_{i,\Delta}(a_{i,\Delta})$ and $b^{\dag}_{i,\Delta}(b_{i,\Delta})$ are the
creation(annihilation) operators for the positive and negative
energy solutions(particle and antiparticle), where $\Delta $ denotes
$M,I,II$. A combination of Minkowski mode, called Unruh mode, can be
transformed into monochromatic Rindler mode and shares the same
vacuum. It holds the relation
\begin{equation}
A_{\Omega,R/L}\equiv cos r_{\Omega}a_{\Omega,I/II} - sin r_{\Omega} b^{\dag}_{\Omega,II/I} \nonumber\\
\end{equation}
Here $cos \gamma_{\Omega}=(e^{\frac{-2 \pi \Omega c}{a}}+1)^{-1/2}$.
Actually we may obtain more general relation such as
\begin{equation} a^{\dag}_{\Omega,U}=q_{L}(A^{\dag}_{\Omega ,L}\otimes
I_{R}) + q_{R}(I_{L} \otimes  A^{\dag}_{\Omega ,R})
\end{equation}
, which goes beyond the single mode approximation. Using this
relation, in case of Grassmann scalar, the Unruh vacuum can be given
by
\begin{eqnarray}
|0_{\Omega }\rangle_{U} &=& cos^{2} \gamma_{\Omega } |0000\rangle_{\Omega
} - sin \gamma_{\Omega }cos \gamma_{\Omega } |0011\rangle_{\Omega }
\nonumber\\
        &+& sin \gamma_{\Omega }cos \gamma_{\Omega } |1100\rangle_{\Omega } - sin^{2}
\gamma_{\Omega } |1111\rangle_{\Omega }
\end{eqnarray}
Here we use the notation $|pqmn\rangle_{\Omega } \equiv |p_{\Omega
}\rangle^{+}_{I}|q_{\Omega }\rangle^{-}_{II}  |m_{\Omega
}\rangle^{-}_{I} |n_{\Omega }\rangle^{+}_{II} $. The two different
one-particle states can be obtained as
\begin{eqnarray}
|1^{+}_{\Omega }\rangle_{U} &=& q_{R}(cos \gamma_{\Omega }
|1000\rangle_{\Omega } - sin \gamma_{\Omega } |1011\rangle_{\Omega
})
\nonumber\\
                        &+& q_{L}(sin \gamma_{\Omega } |1101\rangle_{\Omega } +cos
 \gamma_{\Omega } |0001\rangle_{\Omega } \nonumber\\
 |1^{-}_{\Omega }\rangle_{U} &=& q_{L}(cos \gamma_{\Omega }
|0100\rangle_{\Omega } - sin \gamma_{\Omega } |0111\rangle_{\Omega
})
\nonumber\\
                        &+& q_{R}(sin \gamma_{\Omega } |1110\rangle_{\Omega } +cos
 \gamma_{\Omega } |0010\rangle_{\Omega }
\end{eqnarray}
Here we consider $q_{R}$ and $q_{L}$ as real number. From now on,
for simplicity, the index $\Omega$ will be omitted.
 It should be noted that the physical ordering of the fermionic system was recently introduced by \cite{ref:montero2}\cite{ref:montero3}\cite{ref:chang}.
 So in this report we will use the
physical structure proposed by
\cite{ref:montero1}\cite{ref:montero2}\cite{ref:chang}.

\section*{III. Classical and quantum channel for fermionic system in accelerated frame }
\subsection*{A. The 2 party generalized entangled state $\Phi^{+}$}
  First of all we consider the generalized entangled states $\Phi^{s}$ state and
  $\Phi^{d}$ for the ingredient of quantum channel.
\begin{eqnarray}
|\Phi^{s}\rangle &=& \cos \alpha |0\rangle_{M}|0\rangle_{U} +sin
\alpha
|1\rangle_{M}|1^{+}\rangle_{U} \\
 |\Phi^{d}\rangle&=& \cos \alpha
|1^{+}\rangle_{M}|1^{+}\rangle_{U} +sin \alpha
|1^{-}\rangle_{M}|1^{-}\rangle_{U}
\end{eqnarray}
First of all, it should be noted that eq(7) can be considered as a
fermionic analog of dual rail encoding for bosonic system. Here two
parties Alice and Bob prepare generalized entangled states such as
$\Phi^{s}$ and $\Phi^{d}$ in inertial frames and afterward Bob moves
in a uniformly accelerated frame. Here we assume that Bob and
anti-Bob's detector cannot distinguish between his particle and
antiparticle. Since Bob has inaccessible part due to his
acceleration, when we go beyond the single mode approximation, the
states that Alice and Bob(in Bob's region I)
 may share can be found by tracing out the unaccessible part(Bob's
 region II), which we denote as $\rho^{\Phi^{s}}_{AB_{I}}$ and $\rho^{\Phi^{d}}_{AB_{I}}$.
 In the same way since Bob has inaccessible part due to his
acceleration, when we go beyond the single mode approximation, the
state that Alice and antiBob(in Bob's region II)
 may share can be obtained by tracing out the Bob's region I. Here we denote the
state that Alice and antiBob(in Bob's region II) share as
$\rho^{\Phi^{s}}_{AB_{II}}$ and $\rho^{\Phi^{d}}_{AB_{II}}$.

\begin{figure}[h!]
\begin{center}  
  \includegraphics[width=7cm]{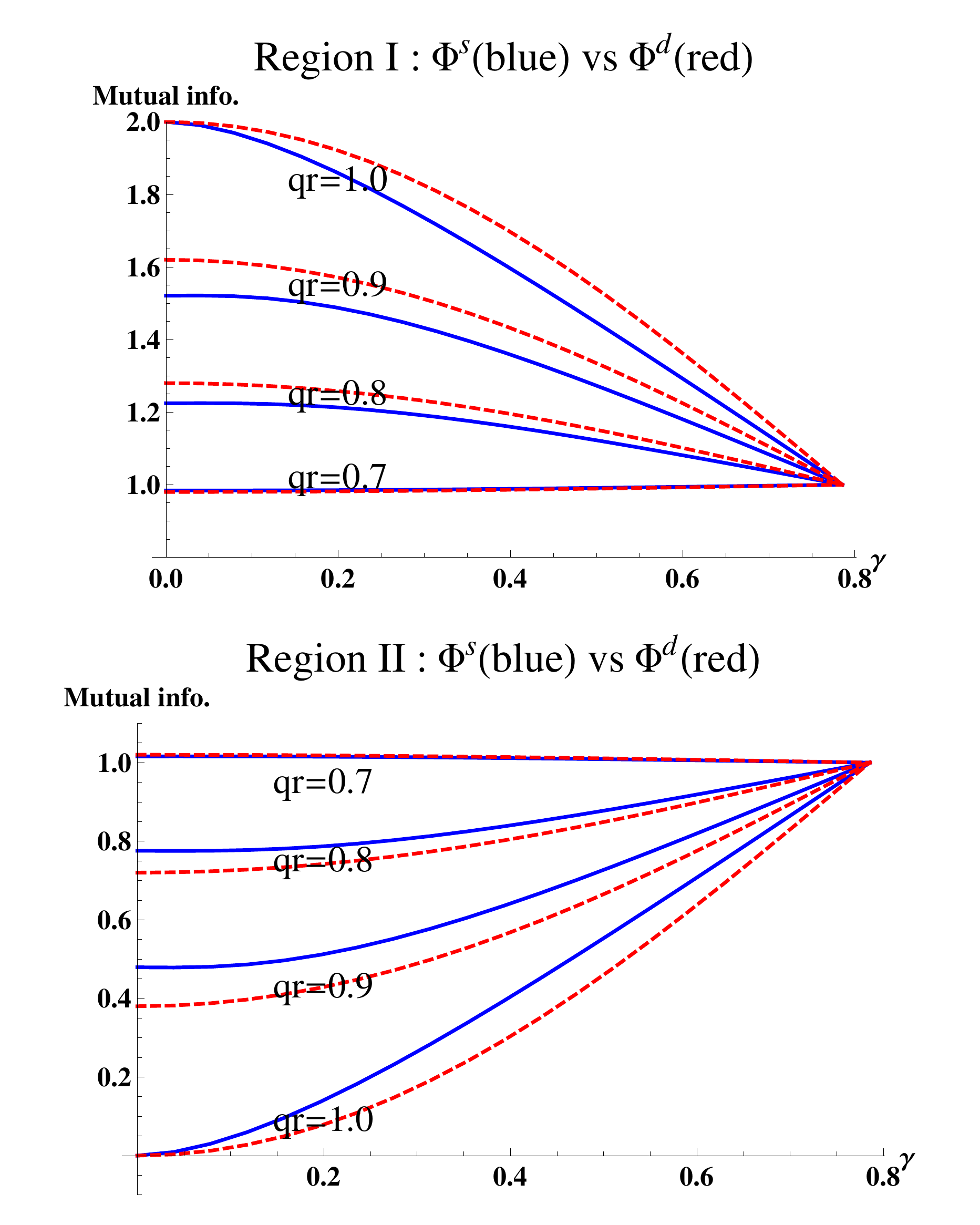}
  \caption{\label{fig2} The mutual information of $\rho^{\Phi^{s}}_{AB_{I}}$,$\rho^{\Phi^{s}}_{AB_{II}}$,$\rho^{\Phi^{d}}_{AB_{I}}$ and $\rho^{\Phi^{d}}_{AB_{II}}$.
  Part (a)((b)) shows the case of $\rho^{\Phi^{s}}_{AB_{I}}$ and $\rho^{\Phi^{d}}_{AB_{I}}$($\rho^{\Phi^{s}}_{AB_{II}}$ and $\rho^{\Phi^{d}}_{AB_{II}}$).
  The solid(dotted) lines denote the mutual information of $
\rho^{\Phi^{s}}_{AB_{I}}$ and $ \rho^{\Phi^{s}}_{AB_{II}}$($
\rho^{\Phi^{d}}_{AB_{I}}$ and $ \rho^{\Phi^{d}}_{AB_{II}}$). In part
(a)the lines from top to bottom denote the mutual information of
$q_{R}=1$, $q_{R}=0.9$, $q_{R}=0.8$ and $q_{R}=\frac{1}{\sqrt{2}}$
respectively, when $\alpha=\frac{\pi}{4}$. In part (b)the lines from
bottom to top denote the mutual information of $q_{R}=1$,
$q_{R}=0.9$, $q_{R}=0.8$ and $q_{R}=\frac{1}{\sqrt{2}}$
respectively, when $\alpha=\frac{\pi}{4}$. $\gamma=\frac{\pi}{4}$
means the infinite acceleration. As it can be seen, the mutual
information of $\rho^{\Phi^{s}}_{AB_{I}}$ and $
\rho^{\Phi^{d}}_{AB_{I}}$($\rho^{\Phi^{s}}_{AB_{II}}$ and $
\rho^{\Phi^{d}}_{AB_{II}}$) coincides at infinite acceleration. }
\end{center}
\end{figure}

 We can consider the classical channels such as
\begin{eqnarray}
\rho^{s}&=&cos\alpha |00\rangle\langle 00| + sin\alpha
|11^{+}\rangle\langle 11^{+}|\\
\rho^{d}&=&cos\alpha |1^{+}1^{+}\rangle\langle 1^{+}1^{+}| +sin
\alpha |1^{-}1^{-}\rangle\langle 1^{-}1^{-}|
\end{eqnarray}
Here $\rho_{d}$ can be interpreted as fermionic version for
classical channel in dual rail encoding. Two parties Alice and Bob
prepare the classical channels such as $\rho^{s}$ and $\rho^{d}$ in
inertial frames and afterward, Bob moves in a uniformly accelerated
frame. The classical channels between Alice and Bob(Bob's region
I)(or between Alice and antiBob(Bob's region II)) can be obtained by
tracing out the part of Bob's region II(that of Bob's region I).
That is, since Bob has inaccessible part due to his acceleration,
when we go beyond the single mode approximation, the classical
channels that Alice and Bob(in Bob's region I) may share are
${\rho^{s}}_{AB_{I}}$ and ${\rho^{d}}_{AB_{I}}$. Likewise, the
classical channels that Alice and antiBob(in Bob's region II) may
share become
${\rho^{s}}_{AB_{II}}$ and ${\rho^{d}}_{AB_{II}}$.\\
The channel capacity between two party $A,B$ can be found from the
mutual information
\begin{equation}
S(\rho_{A}:\rho_{B})=S(\rho_{A})+S(\rho_{B})-S(\rho_{AB})
\end{equation}
Here $S(\rho)$ denotes the Von Neumann entropy given by $-tr(\rho
log\rho)$. The channel capacity can be obtained by maximum value for
the mutual information. Actually, the channel capacity will be
useful for evaluating the classical channel. The effectiveness of
the quantum channel can be measured by the conditional entropy given
by

\begin{equation}
S(\rho_{A}|\rho_{B})=S(\rho_{AB})-S(\rho_{B})
\end{equation}

The mutual information of $\rho^{\Phi^{s}}_{AB_{I}}$,
$\rho^{\Phi^{s}}_{AB_{II}}$, $\rho^{\Phi^{d}}_{AB_{I}}$ and
$\rho^{\Phi^{d}}_{AB_{II}}$ can be found in Fig 1. The behavior of
the mutual information for those quantum states is studied in terms
of $\gamma$ and $q_{R}$. At first, the mutual information for Alice
and Bob when $q_{R}=1$ or $q_{R}=0.9$ or $q_{R}=0.8$, decreases as
the acceleration increases. And the degree for decrease is
different. Furthermore, it is worthwhile to note that the mutual
information for $ \rho^{\Phi^{d}}_{AB_{I}}$ is greater than that of
$ \rho^{\Phi^{s}}_{AB_{I}}$ as the acceleration increases. Actually,
in an inertial frame there are four different maximally entangled
states, but they are locally equivalent in terms of local unitary
transformations. However, in an accelerated frame we may have
different entangled states such as $ \Phi^{s}$ and $\Phi^{d}$, which
shows different mutual information. Also it should be addressed that
the mutual information of each state has the same value at the
infinite acceleration regardless of $q_{R}$. Part (b) in Fig.1.
depicts the mutual information of $ \rho^{\Phi^{s}}_{AB_{II}}$ and $
\rho^{\Phi^{d}}_{AB_{II}}$. In part (b) the mutual information for
Alice and antiBob when $q_{R}=1$ or $q_{R}=0.9$ or $q_{R}=0.8$,
increases as the acceleration increases.\\

\begin{figure}[h!]
\begin{center}  
  \includegraphics[width=7cm]{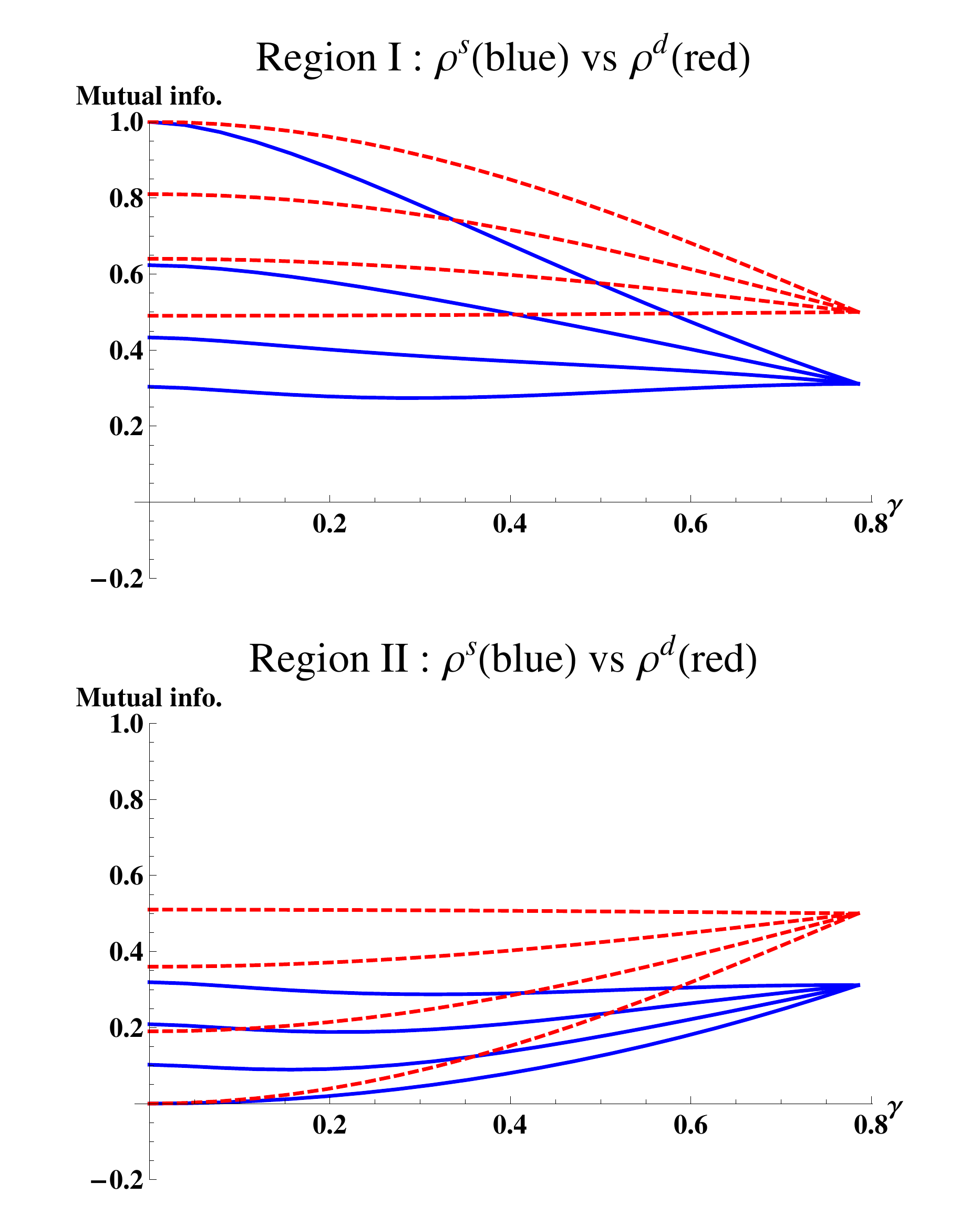}
  \caption{\label{fig2} The mutual information of classical channels ${\rho^{s}}_{AB_{I}}$,${\rho^{d}}_{AB_{I}}$,${\rho^{s}}_{AB_{II}}$ and ${\rho^{d}}_{AB_{II}}$.
  Part (a)((b)) shows the case of ${\rho^{s}}_{AB_{I}}$ and ${\rho^{d}}_{AB_{I}}$(${\rho^{s}}_{AB_{II}}$ and ${\rho^{d}}_{AB_{II}}$).
  The solid(dotted) lines denote the mutual information of ${\rho^{s}}_{AB_{I}}$ and ${\rho^{d}}_{AB_{I}}$(${\rho^{s}}_{AB_{II}}$ and ${\rho^{d}}_{AB_{II}}$). In part
(a)the lines from top to bottom denote the mutual information of
$q_{R}=1$, $q_{R}=0.9$, $q_{R}=0.8$ and $q_{R}=\frac{1}{\sqrt{2}}$
respectively, when $\alpha=\frac{\pi}{4}$. In part (b)the lines from
bottom to top denote the mutual information of $q_{R}=1$,
$q_{R}=0.9$, $q_{R}=0.8$ and $q_{R}=\frac{1}{\sqrt{2}}$
respectively, when $\alpha=\frac{\pi}{4}$. $\gamma=\frac{\pi}{4}$
means the infinite acceleration. }
\end{center}
\end{figure}

The mutual information for classical channel can be found in Fig 2.
As we saw in part (a) in Fig.2, the mutual information of both
classical channels for single and dual rail one between Alice and
Bob when $q_{R}=1$ or $q_{R}=0.9$ may decrease as the acceleration
increases. And it should be noted that the behavior of mutual
information for $\rho^{s}$ is different from that of $\rho^{d}$.
Also smaller $q_{R}$ is lesser the mutual information of both
classical channels is. At each $q_{R}$ the mutual information of
$\rho^{d}$ is larger than that of $\rho^{s}$. Part (b) in Fig.2
denote the mutual information of both classical channels for single
and dual rail one between Alice and antiBob.

\begin{figure}[h!]
\begin{center}  
  \includegraphics[width=8.5cm]{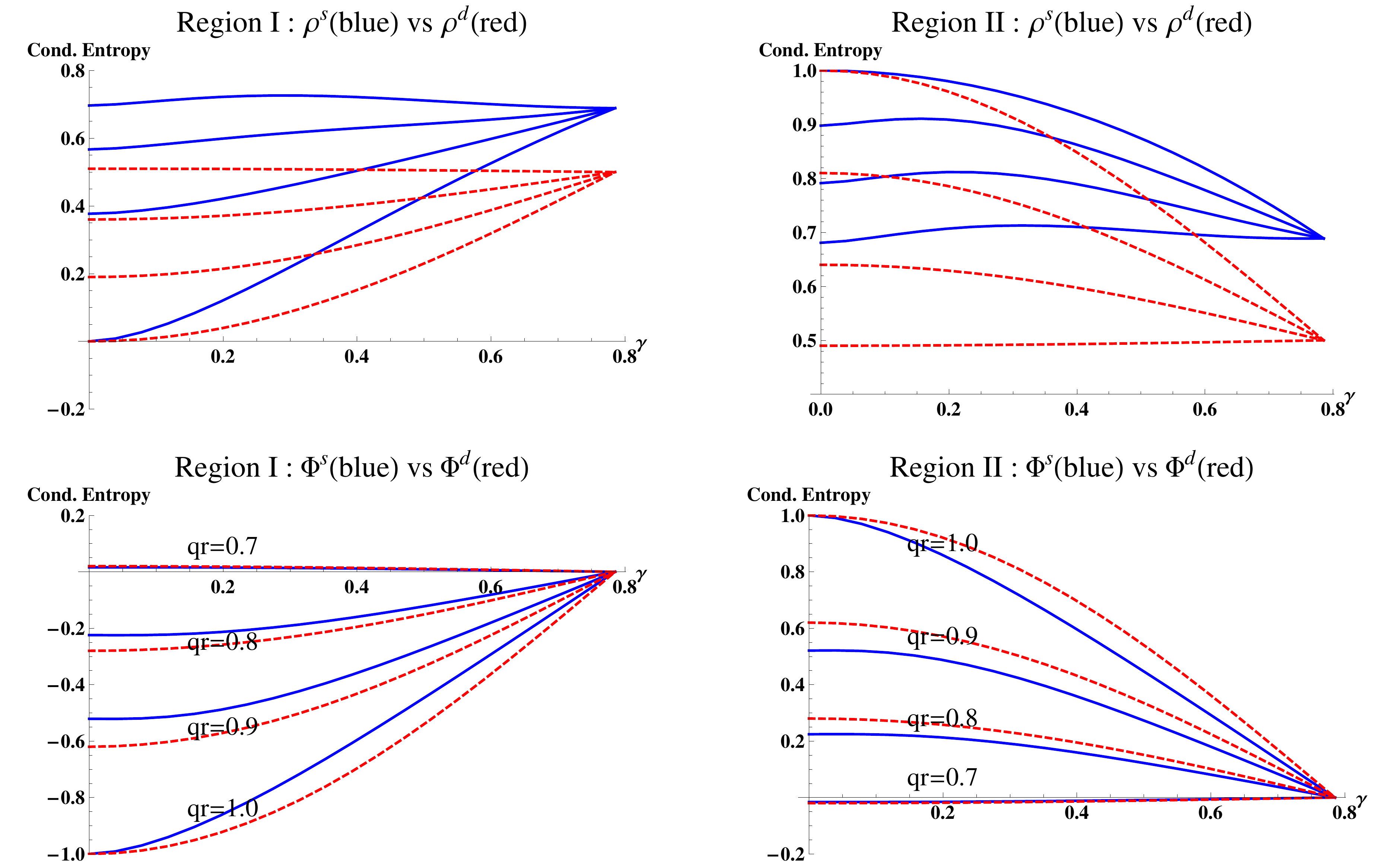}
  \caption{\label{fig2} The conditional entropy of ${\rho^{s}}_{AB_{I}}$,${\rho^{d}}_{AB_{I}}$,${\rho^{s}}_{AB_{II}}$,${\rho^{d}}_{AB_{II}}$,$\rho^{\Phi^{s}}_{AB_{I}}$,$\rho^{\Phi^{d}}_{AB_{I}}$,$\rho^{\Phi^{s}}_{AB_{II}}$ and $\rho^{\Phi^{d}}_{AB_{II}}$.
  Part (a),(b),(c), and (d) denote the conditional entropy of ${\rho^{s}}_{AB_{I}}$ and ${\rho^{d}}_{AB_{I}}$,${\rho^{s}}_{AB_{II}}$ and ${\rho^{d}}_{AB_{II}}$,$\rho^{\Phi^{s}}_{AB_{I}}$ and $\rho^{\Phi^{s}}_{AB_{II}}$, and $\rho^{\Phi^{d}}_{AB_{I}}$ and $\rho^{\Phi^{d}}_{AB_{II}}$ respectively.
  $\gamma=\frac{\pi}{4}$ means the infinite acceleration.
  As it can be seen, the conditional entropy of $\rho^{\Phi^{s}}_{AB_{I}}$ and $\rho^{\Phi^{d}}_{AB_{I}}$($\rho^{\Phi^{s}}_{AB_{II}}$ and $\rho^{\Phi^{d}}_{AB_{II}}$) coincides at infinite acceleration. }
\end{center}
\end{figure}

The conditional entropy of
${\rho^{s}}_{AB_{I}}$,${\rho^{d}}_{AB_{I}}$,${\rho^{s}}_{AB_{II}}$,${\rho^{d}}_{AB_{II}}$,$\rho^{\Phi^{s}}_{AB_{I}}$,$\rho^{\Phi^{d}}_{AB_{I}}$,$\rho^{\Phi^{s}}_{AB_{II}}$
and $\rho^{\Phi^{d}}_{AB_{II}}$ can be found in Fig.3. Part
(a),(b),(c), and (d) in Fig.3 denote the conditional entropy of
${\rho^{s}}_{AB_{I}}$ and
${\rho^{d}}_{AB_{I}}$,${\rho^{s}}_{AB_{II}}$ and
${\rho^{d}}_{AB_{II}}$,$\rho^{\Phi^{s}}_{AB_{I}}$ and
$\rho^{\Phi^{s}}_{AB_{II}}$, and $\rho^{\Phi^{d}}_{AB_{I}}$ and
$\rho^{\Phi^{d}}_{AB_{II}}$ respectively.
  As it can be seen in part (a) and (b) in Fig.3, the conditional entropy of ${\rho^{s}}_{AB_{I}}$ and ${\rho^{d}}_{AB_{I}}$(${\rho^{s}}_{AB_{II}}$ and
${\rho^{d}}_{AB_{II}}$) always possesses positive value. However,
part (c) in Fig.3 clearly shows the negative conditional entropy of
$\rho^{\Phi^{s}}_{AB_{I}}$ and $\rho^{\Phi^{d}}_{AB_{I}}$ when the
acceleration has finite value. In fact, the conditional entropy can
be nicely understood in terms of the strong additivity. The strong
additivity can be found from eq(12) such as

\begin{equation}
A(\rho_{A},\rho_{B},\rho_{C})=S(\rho_{AB})-S(\rho_{A})+S(\rho_{AC})-S(\rho_{C})
\geq 0
\end{equation}

In order to investigate the strong additivity we consider the Werner
state, where a white noise is added to a maximally entangled states.
The mixedness of Werner states is parameterized by a single
parameter. So we suppose that two parties Alice and Bob prepare
Werner states in inertial frames, and then Bob moves in the
uniformly accelerated frame. That is, the initial state of Alice and
Bob can be expressed as follows,
\begin{equation}
\rho_{W}= F |\Phi^{s} (\alpha=\pi/4) \rangle  \langle \Phi^{s}
(\alpha=\pi/4) | + \frac{1-F}{4}\mathbb{I}\label{eq9},
\end{equation}
where the maximally entangled state is taken from Eq.8 when
$\alpha=\pi/4$. Suppose that Bob moves in an accelerated frame.
Beyond the single-mode approximation, the state that Alice and
Bob(antiBob) share in Bob's region $I$($II$) is obtained by tracing
the region $II$($I$) And we denote the quantum state as
${\rho^{W}}_{AB_{I}}$(${\rho^{W}}_{AB_{II}}$).

\begin{figure}[h!]
\begin{center}  
  \includegraphics[width=8.5cm]{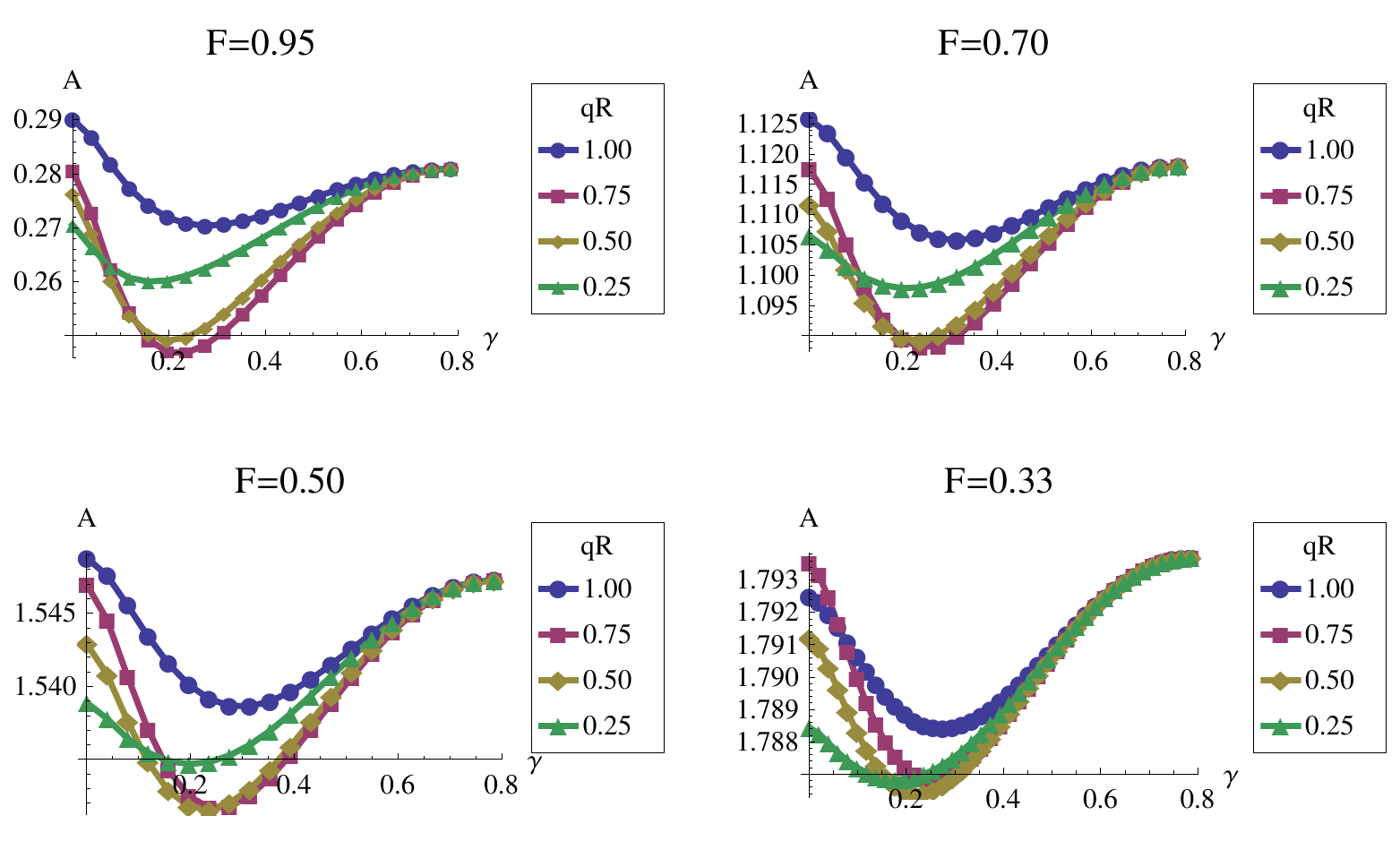}
  \caption{\label{fig2} The strong additivity of ${\rho^{W}}_{AB_{I}}$. Part (a),(b),(c) and (d) of Fig. 3 depict the
inequality for conditional entropy of Werner state with
$F=0.95,F=0.70,F=0.50$,$F=0.33$ respectively. The circle, rectangle,
diamond, and triangle denote the strong additivity of
${\rho^{W}}_{AB_{I}}$ at $q_{R}=1$, $q_{R}=0.75$, $q_{R}=0.5$ and
$q_{R}=0.25$ respectively. $\gamma=\frac{\pi}{4}$ means the infinite
acceleration. }
\end{center}
\end{figure}

The strong additivity for Werner state can be seen in Fig.4. It
explains the behavior of strong additivity of the state
${\rho^{W}}_{AB_{I}}$. Part (a),(b),(c) and (d) of Fig. 4 depict the
inequality for conditional entropy of Werner state with
$F=0.95,F=0.70,F=0.50$,$F=0.33$ respectively. Specially Part (d)
shows why we need a method beyond single mode approximation, where
the value at $q_{R}=1$ is lesser than that of $q_{R}=0.75$ in some
region of acceleration.

\section*{IV. Discussion and Conclusion}
In this article, we investigated the communication channel of
fermionic system in an accelerated frame. We considered classical
and quantum channels in terms of the single and dual rail encoding.
We found that at the infinite acceleration, the mutual information
of single rail quantum channel coincides with that of double rail
quantum channel, but those of classical ones reach different values.
In view of conditional entropy, we observed, that at the infinite
acceleration, the conditional entropy of single(or double) rail
quantum channel vanishes, even though those of classical ones may
have finite values. Furthermore, we saw that even when considering a
method beyond the single mode approximation, the dual rail entangled
state seems to provide better channel capacity than another
maximally entangled state, unlike the bosonic case recently studied
by Montero and Mart\'{i}n-Mart\'{i}nez. Moreover, we found that the
single-mode approximation is not sufficient to analyze the
communication channel of fermionic system in an accelerated frame.

\section*{Acknowledgement}
 We would like to thank Dr.Mart\'{i}n-Mart\'{i}nez for valuable comments. This work is supported by Basic Science Research Program through the National Research
Foundation of Korea funded by the Ministry of Education, Science and
Technology (KRF2011-0027142).

\end{document}